\documentclass[aps,prb,reprint,longbibliography,noeprint,superscriptaddress]{revtex4-2}
\usepackage{amsfonts,amssymb,amsmath,bm,graphicx,xcolor}
\usepackage[colorlinks=true,citecolor=blue,linkcolor=blue,urlcolor=blue]{hyperref}
\allowdisplaybreaks[4]
\DeclareMathOperator{\re}{Re}
\DeclareMathOperator{\Tr}{Tr}
\DeclareMathOperator{\tr}{tr}
\newcommand{\complex}{\mathrm{c.c.}}
\newcommand{\ham}{(\mathrm{h})} 
\newcommand{\wan}{(\mathrm{w})} 
\begin{document}
\title{Wannier interpolation of spin accumulation coefficient}
\author{Atsuo Shitade}
\email{shitade@sanken.osaka-u.ac.jp}
\affiliation{Institute of Scientific and Industrial Research, The University of Osaka, Ibaraki, Osaka, Japan}
\author{Emi Minamitani}
\affiliation{Institute of Scientific and Industrial Research, The University of Osaka, Ibaraki, Osaka, Japan}
\begin{abstract}
  The spin Hall (SH) effect is widely understood as a phenomenon in which spin current flows perpendicular to an electric field.
  In the presence of a spin-orbit coupling, however, spin current is ambiguous,
  and the SH conductivity depends on the definition of spin current.
  In this article, we develop an \textit{ab initio} computational scheme for the spin accumulation coefficient,
  which characterizes the spin accumulation and would be an alternative indicator of the SH effect.
  The proposed method has been implemented into an open-source software Wannier90
  and serves high-precision \textit{ab initio} research on the SH effect.
\end{abstract}
\maketitle
Spintronics is a research field to exploit the spin degree of freedom of electrons
for next-generation devices with high-speed processing and low energy consumption.
Spin can be generated electrically even in nonmagnetic systems via spin-orbit couplings (SOCs).
One of such phenomena is the spin Hall (SH) effect
in which spin current flows perpendicular to an applied electric field
and turns into the spin accumulation at the surface~\cite{RevModPhys.87.1213}.
Since the experimental observations in semiconductors~\cite{Kato1910,PhysRevLett.94.047204},
the reseach on the SH effect has been extending to various materials with large SOCs
such as heavy metals~\cite{RevModPhys.87.1213}, transition metal dichalcogenides (TMDCs)~\cite{SHI2023100060},
and antiferromagnets~\cite{RevModPhys.90.015005}.

The SH conductivity, namely, the response of spin current to an electric field,
has been widely evaluated as an indicator of the SH effect.
However, in the presence of SOCs, spin is not conserved, and spin current is ambiguous.
The conventional spin current $\hat{J}_{sa}^{\phantom{sa} i} = \{\hat{s}_{a}, \hat{v}^{i}\} / 2$,
in which $\hat{s}_{a} = (\hbar / 2) \sigma_{a}$ and $\hat{v}^{i}$ are
the spin operator with the Pauli matrix $\sigma_{a}$ and velocity operator, respectively,
has been chosen in most of the literature~\cite{PhysRevLett.94.226601,PhysRevLett.95.156601}
but suffers some critical problems such as the equilibrium spin current and the absence of a conjugate force,
as discussed more in Sec.~\ref{sec:discussion}.
Instead, we may choose the conserved spin current consisting of the conventional one and spin torque dipole moment~\cite{PhysRevLett.96.076604},
leading to a different value of the SH conductivity~\cite{Ma2024}.
There is no guiding principle what definition of spin current should be chosen.

Spin is well defined in contrast to spin current,
and the spin accumulation at the surface has been experimentally observed~\cite{Kato1910,PhysRevLett.94.047204}.
For the Rashba model with nonmagnetic disorder, the spatial distribution of the spin density was numerically computed
by solving the coupled diffusion equations~\cite{PhysRevLett.93.226602,PhysRevB.70.195343,RASHBA200631,PhysRevB.74.035340}
or the Landauer-Keldysh formalism~\cite{PhysRevLett.95.046601,PhysRevB.72.081301,PhysRevB.73.075303}.
In this system, the spin accumulation does occur, while the SH conductivities of both the conventional~\cite{%
PhysRevLett.93.226602,PhysRevB.70.041303,PhysRevB.71.033311,PhysRevB.71.245318,PhysRevB.73.049901,PhysRevB.71.245327,PhysRevLett.96.056602}
and conserved spin current~\cite{PhysRevB.73.113305} vanish.
These results indicate that spin, rather than spin current, is the primary object.
However, it is difficult to compute the spin accumulation for real materials
because we need to impose the open boundary conditions or attach the leads, apply a voltage, and deal with disorder.

In this article, we develop an \textit{ab initio} computational scheme for the spin accumulation coefficient (SAC),
namely, the response of spin to an electric field gradient~\cite{PhysRevB.98.174422,PhysRevB.105.L201202,PhysRevB.106.045203}.
The SAC characterizes the spin accumulation at the surface owing to the SH effect but, counterintuitively,
can be evaluated as a bulk property using Bloch wavefunctions.
Hence, the SAC would be an alternative indicator of the SH effect.
With the help of maximally localized Wannier functions~\cite{RevModPhys.84.1419} implemented in an open-source software Wannier90~\cite{Pizzi_2020},
we can evaluate the SAC with high precision for real materials.
We apply our method to monolayer TMDC MoS$_{2}$ and trigonal tellurium
to confirm the consistency with the point group symmetry and the gauge invariance regarding Wannier functions.
The SAC is not correlated with the SH conductivity and free from the critical problems of spin current.
Our work contributes to quantitative materials research on the SH effect based on solid foundations.

\section{Results}\label{sec:results}
\subsection{Wannier interpolation of SAC}\label{sub:wannier}
We consider the response of spin to an electric field gradient,
$\langle \Delta \hat{s}_{a} \rangle = g_{sa}^{\phantom{sa} ij} \partial_{x^{i}} E_{j}$.
Within the relaxation time approximation, the SAC can be evaluated using Bloch wavefunctions as
$g_{sa}^{\phantom{sa} ij} = -\tau \gamma_{sa}^{\phantom{sa} ij}$ with~\cite{PhysRevB.105.L201202}
\begin{align}
    \gamma_{sa}^{\phantom{sa} ij}
    = & -\frac{e}{\hbar} \sum_{n} \int \frac{d^{3} k}{(2 \pi)^{3}} [-f^{\prime}(\epsilon_{n}(\bm{k}))] \notag \\
    & \times [s_{na}^{\phantom{na} i}(\bm{k}) \partial_{k_{j}} \epsilon_{n}(\bm{k}) - s_{na}(\bm{k}) \epsilon^{ijk} m_{nk}(\bm{k})],
    \label{eq:spin_accumulation}
\end{align}
in which $\tau$ is the phenomenological relaxation time, $-e$ is the elementary charge,
and $f(\epsilon) = [e^{(\epsilon - \mu) / k_{\mathrm{B}} T} + 1]^{-1}$ is the Fermi distribution function
at the chemical potential $\mu$ and temperature $T$.
\begin{subequations} \begin{align}
    s_{na}^{\phantom{na} i}(\bm{k})
    = & \frac{i}{2} \langle u_{n}(\bm{k}) | \hat{s}_{a} \hat{Q}_{n}(\bm{k}) | \partial_{k_{i}} u_{n}(\bm{k}) \rangle + \complex, \label{eq:spin_magnetic_quadrupole} \\
    s_{na}(\bm{k})
    = & \langle u_{n}(\bm{k}) | \hat{s}_{a} | u_{n}(\bm{k}) \rangle, \label{eq:spin} \\
\epsilon^{ijk} m_{nk}(\bm{k})
    = & \frac{-i}{2}
    \langle \partial_{k_{i}} u_{n}(\bm{k}) | [\epsilon_{n}(\bm{k}) - \hat{H}(\bm{k})] | \partial_{k_{j}} u_{n}(\bm{k}) \rangle \notag \\
    & + \complex, \label{eq:orbital_magnetic}
\end{align} \label{eq:geometric}\end{subequations}
are the spin magnetic quadrupole moment~\cite{PhysRevB.99.024404}, spin polarization,
and orbital magnetic moment~\cite{PhysRevLett.95.137204,PhysRevLett.95.169903}, respectively,
in which $\hat{Q}_{n}(\bm{k}) = 1 - | u_{n}(\bm{k}) \rangle \langle u_{n}(\bm{k}) |$,
and $\epsilon_{n}(\bm{k})$ and $| u_{n}(\bm{k}) \rangle$ are the eigenvalues and eigenstates of the Bloch Hamiltonian $\hat{H}(\bm{k})$.
$\gamma_{sa}^{\phantom{sa} ij}$ has the same tensor structure and dimension as the SH conductivity for any point group.
Equation~\eqref{eq:spin_accumulation} is the only term in nonmagnetic systems,
while the Fermi-sea term is allowed in magnetic systems as mentioned in Sec.~\ref{sec:discussion}.

To evaluate Eq.~\eqref{eq:spin_magnetic_quadrupole} in the scheme of Wannier functions,
we introduce the trace formula of the spin magnetic quadrupole moment as
\begin{equation}
    Q_{a}^{\phantom{a} i}(\bm{k})
    = i \Tr [\hat{s}_{a} \hat{Q}(\bm{k}) \partial_{k_{i}} \hat{P}(\bm{k})]. \label{eq:trace_q}
\end{equation}
Here, $\hat{P}(\bm{k}) = | u(\bm{k}) \rangle f(\bm{k}) \langle u(\bm{k}) |$ is the projection operator to the occupied subspace,
and $\hat{Q}(\bm{k}) = 1 - \hat{P}(\bm{k})$.
We have two gauge choices for a set of Bloch wavefunctions $| u(\bm{k}) \rangle$.
One is the Wannier gauge that is the Fourier transform of Wannier functions, denoted by $| u^{\wan}(\bm{k}) \rangle$.
The other is the Hamiltonian gauge that diagonalizes
$\mathbb{H}^{\wan}(\bm{k}) = \langle u^{\wan}(\bm{k}) | \hat{H}(\bm{k}) | u^{\wan}(\bm{k}) \rangle$,
denoted by $| u^{\ham}(\bm{k}) \rangle$.
Hereafter, a gauge choice is not specified unless explicitly shown.
In the Hamiltonian gauge, $f(\bm{k})$ is a diagonal matrix with $f_{n}^{\ham}(\bm{k}) = 0, 1$.
We also define the projection operator to the Wannier subspace as $\hat{\mathbb{P}}(\bm{k}) = | u(\bm{k}) \rangle \langle u(\bm{k}) |$
and $\hat{\mathbb{Q}}(\bm{k}) = \mathbb{I} - \hat{\mathbb{P}}(\bm{k})$,
which is related by $\hat{Q}(\bm{k}) = \hat{\mathbb{Q}}(\bm{k}) + \hat{Q}_{\mathrm{in}}(\bm{k})$
with $\hat{Q}_{\mathrm{in}}(\bm{k}) = | u(\bm{k}) \rangle g(\bm{k}) \langle u(\bm{k}) |$ ($f(\bm{k}) + g(\bm{k}) = 1$).
Equation~\eqref{eq:trace_q} expressed by uppercase $\Tr$, which means trace over the full Hilbert space, is gauge invariant by construction.

Next, to compute Eq.~\eqref{eq:trace_q} efficiently,
we rewrite Eq.~\eqref{eq:trace_q} using Wannier matrix elements defined only in the Wannier subspace.
Following ref.~\cite{PhysRevB.85.014435} for the Berry curvature and orbital magnetic moment, we obtain
\begin{equation}
    Q_{a}^{\phantom{a} i}(\bm{k})
    = \tr [\tilde{\mathbb{Q}}_{a}^{\phantom{a} i}(\bm{k}) f(\bm{k}) + i \mathbb{S}_{a}(\bm{k}) g(\bm{k}) \tilde{\partial}_{k_{i}} f(\bm{k})],
    \label{eq:trace_covariant_q}
\end{equation}
in which
\begin{subequations} \begin{align}
    \tilde{\mathbb{Q}}_{a}^{\phantom{a} i}(\bm{k})
    = & i \langle u(\bm{k}) | \hat{s}_{a} \hat{\mathbb{Q}}(\bm{k}) | \partial_{k_{i}} u(\bm{k}) \rangle, \label{eq:covariant_saa} \\
    \mathbb{S}_{a}(\bm{k})
    = & \langle u(\bm{k}) | \hat{s}_{a} | u(\bm{k}) \rangle, \label{eq:wannier_ss} \\
    \tilde{\partial}_{k_{i}} f(\bm{k})
    = & \partial_{k_{i}} f(\bm{k}) - i [\mathbb{A}^{i}(\bm{k}), f(\bm{k})], \label{eq:covariant_derivative}
\end{align} \label{eq:covariant}\end{subequations}
with $\mathbb{A}^{i}(\bm{k}) = i \langle u(\bm{k}) | \partial_{k_{i}} u(\bm{k}) \rangle$.
These quantities are gauge covariant, namely, transform as $X^{\ham}(\bm{k}) = U^{\dag}(\bm{k}) X^{\wan}(\bm{k}) U(\bm{k})$
under a gauge transformation $| u^{\ham}(\bm{k}) \rangle = | u^{\wan}(\bm{k}) \rangle U(\bm{k})$.
As a result, Eq.~\eqref{eq:trace_covariant_q} expressed by lowercase $\tr$, which means trace over the Wannier subspace, is also gauge invariant.

Finally, we arrive at
\begin{align}
    Q_{a}^{\phantom{a} i}(\bm{k})
    = & \tr [\mathbb{Q}_{a}^{\phantom{a} i}(\bm{k}) f(\bm{k}) - \mathbb{S}_{a} f(\bm{k}) \mathbb{A}^{i}(\bm{k}) f(\bm{k}) \notag \\
    & + i\mathbb{S}_{a}(\bm{k}) g(\bm{k}) \partial_{k_{i}} f(\bm{k})], \label{eq:trace_wannier_q}
\end{align}
using
$\tilde{\mathbb{Q}}_{a}^{\phantom{a} i}(\bm{k}) = \mathbb{Q}_{a}^{\phantom{a} i}(\bm{k}) - \mathbb{S}_{a}(\bm{k}) \mathbb{A}^{i}(\bm{k})$.
Here, $\mathbb{Q}_{a}^{\phantom{a} i}(\bm{k}) = i \langle u(\bm{k}) | \hat{s}_{a} | \partial_{k_{i}} u(\bm{k}) \rangle$
as well as $\mathbb{S}_{a}(\bm{k})$ and $\mathbb{A}^{i}(\bm{k})$ is computed in Wannier90 from \textit{ab initio} calculations.
In the Hamiltonian gauge, $\partial_{k_{i}} f^{\ham}(\bm{k})$ in Eq.~\eqref{eq:trace_wannier_q} is assumed to be zero, and then we find
\begin{equation}
    \sum_{n} s_{na}^{\ham i}(\bm{k}) f_{n}^{\ham}(\bm{k})
    = \re Q_{a}^{\phantom{a} i}(\bm{k}). \label{eq:occupied_spin_magnetic_quadrupole}
\end{equation}
The trace formula gives a part of the bulk spin magnetic quadrupole moment~\cite{PhysRevB.99.024404},
which is allowed in magnetoelectric materials without either inversion or time-reversal symmetry.

Following ref.~\cite{PhysRevB.97.035158} for the Berry curvature and orbital magnetic moment,
we obtain Eq.~\eqref{eq:spin_magnetic_quadrupole} for the $n$th band
by choosing fake occupations, namely, $f_{m; n}^{\ham}(\bm{k}) = \delta_{mn}$ for given $n$.
Now we are ready to compute the SAC for real materials.

\subsection{Example: monolayer MoS$_{2}$}\label{sub:MoS2}
Here we apply our Wannier interpolation of the SAC to two different materials.
One example is monolayer TMDC MoS$_{2}$.
TMDCs are layered materials with a chemical formula of $M X_{2}$,
in which $M$ and $X$ are transition metal and chalcogen atoms, respectively.
The layers are weakly coupled to each other by the van der Waals (vdW) interaction.
In particular, monolayer systems have attracted much attention from the viewpoints of spintronics and valleytronics,
and were studied in terms of the SH conductivity~\cite{PhysRevB.86.165108}.

\begin{figure*}
    \centering
    \includegraphics[clip,width=\textwidth]{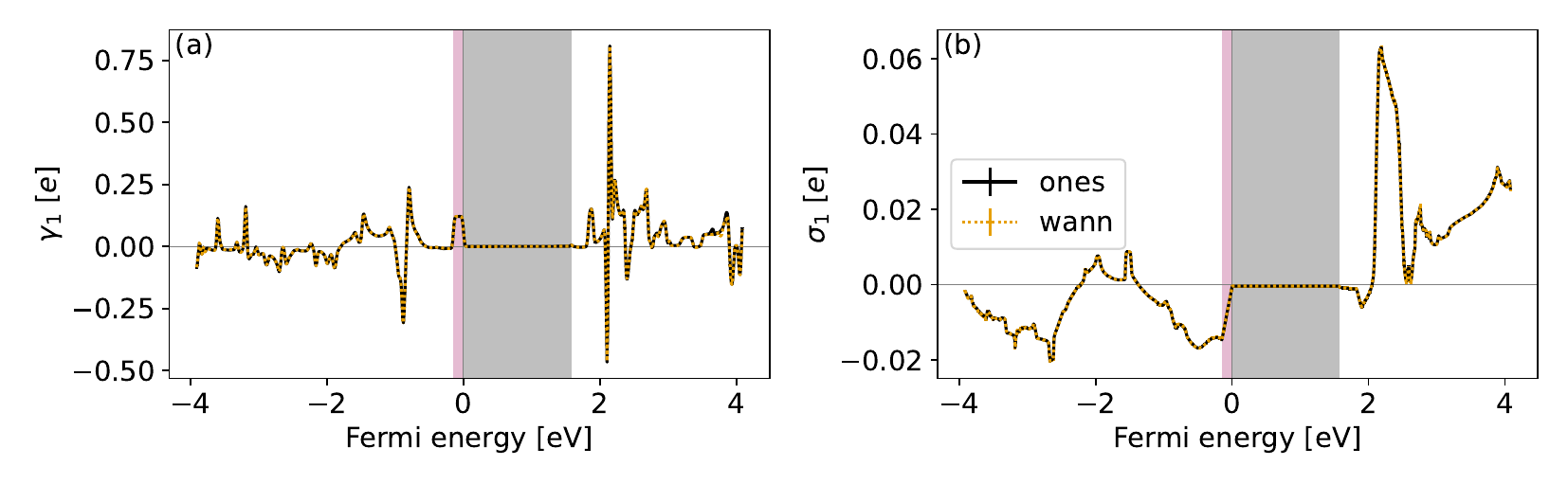}
    \caption{%
    SAC and SH conductivity per layer of monolayer MoS$_{2}$ as a function of the Fermi energy.
    (a) SAC
    $\gamma_{1} = \epsilon^{1a}_{\phantom{1a} ij} \gamma_{sa}^{\phantom{sa} ij} / 2$
    and
    (b) SH conductivity
    $\sigma_{1} = \epsilon^{1a}_{\phantom{a} ij} \sigma_{sa}^{\phantom{sa} ij} / 2$.
    Errors are defined by $\max_{\epsilon_{1} \not= 0} |\gamma_{sa}^{\phantom{sa} ij} - \gamma_{1}|$ etc.
    Black solid and orange dotted lines correspond to the number of iterations for wannierization $= 0$ and $1000$, respectively.
    Gray and purple areas represent the band gap and the spin splitting at the $K$ and $K^{\prime}$ points owing to the Ising SOC, respectively.%
    } \label{fig:Figure1}
\end{figure*}

Figure~\ref{fig:Figure1}a shows the Fermi-energy dependence of the SAC per layer.
See Sec.~\ref{sub:MoS2_details} for computational details.
In this material, the $D_{3h}$ point group symmetry allows the form of
$\gamma_{sa}^{\phantom{sa} ij} = \gamma_{1} \epsilon_{1a}^{\phantom{1a} ij}$,
in which $\epsilon_{1a}^{\phantom{1a} ij}$ is a tensor
whose nonzero components are $\epsilon_{1z}^{\phantom{1z} xy} = -\epsilon_{1z}^{\phantom{1z} yx} = 1$.
The negligible error indicates that the computed SAC is consistent with this symmetry.
We also check the gauge invariance of the SAC by changing the number of iterations for wannierization.
These results validate our implementation.

For comparison, we also show the Fermi-energy dependence of the SH conductivity $\sigma_{sa}^{\phantom{sa} ij}$
of the conventional spin current~\cite{PhysRevB.86.165108} in Fig.~\ref{fig:Figure1}b.
$\sigma_{sa}^{\phantom{sa} ij} = \sigma_{1} \epsilon_{1a}^{\phantom{1a} ij}$ is allowed by the symmetry.
In general, the SAC and SH conductivity are independent of each other.
Right below the Fermi energy painted in purple, where the spin splitting occurs at the $K$ and $K^{\prime}$ points owing to the Ising SOC,
the SAC shows a positive plateau, while the SH conductivity shows negative increase.
Such a plateau in the SAC can be detected in experiments by changing the carrier density.

\subsection{Example: trigonal tellurium}\label{sub:Te}
The other example is trigonal tellurium, in which the nonlinear Hall effect~\cite{PhysRevLett.115.216806}
as well as the orbital~\cite{PhysRevB.92.235205,PhysRevLett.116.077201}
and spin Edelstein effects~\cite{Ivchenko1978,Ivchenko1989,Aronov1989,Edelstein1990233} were studied~\cite{PhysRevB.97.035158}.

\begin{figure*}
    \centering
    \includegraphics[clip,width=\textwidth]{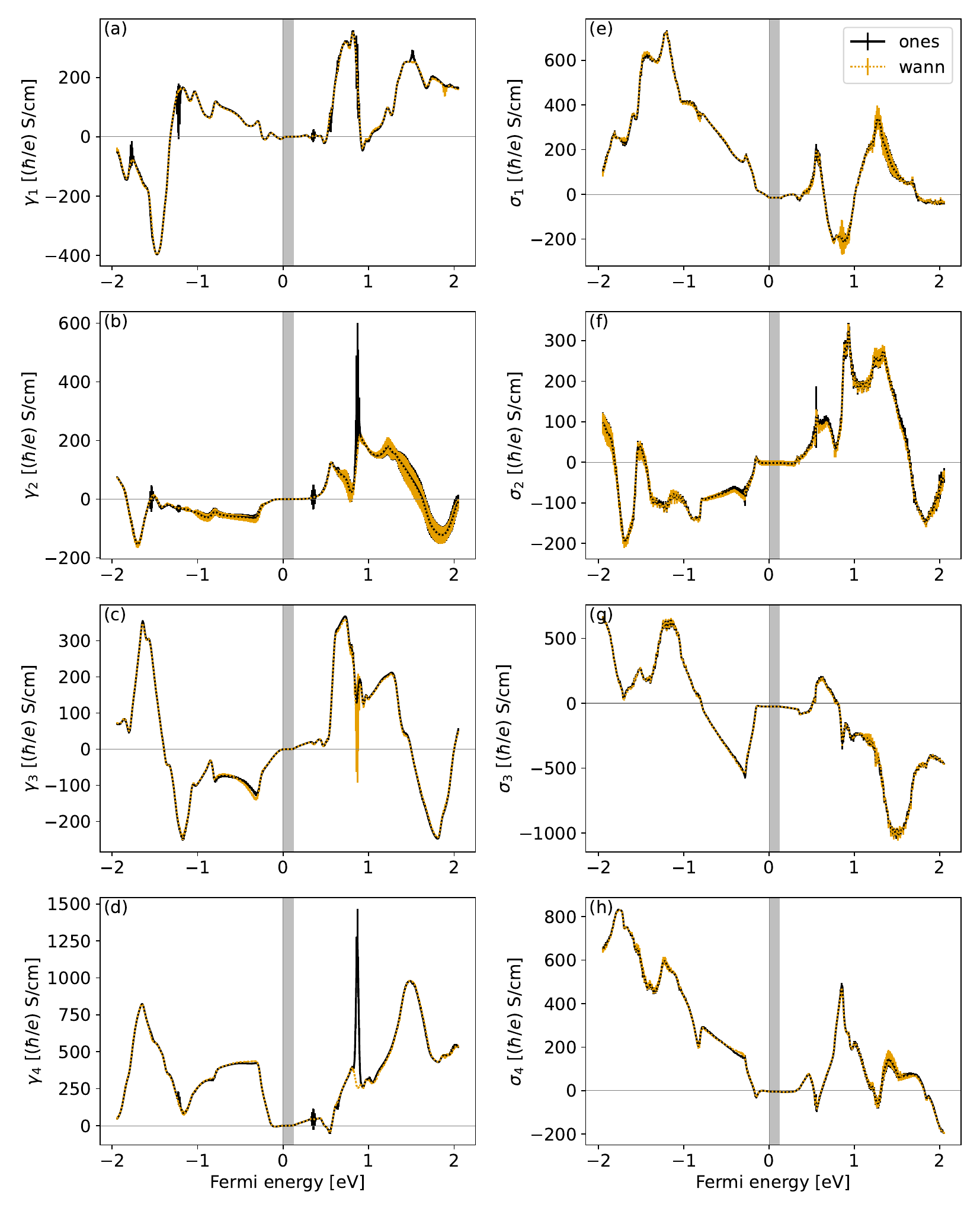}
    \caption{%
    SAC and SH conductivity of trigonal tellurium as a function of the Fermi energy.
    (a-d) SAC
    $\gamma_{\alpha} = \epsilon^{\alpha a}_{\phantom{1a} ij} \gamma_{sa}^{\phantom{sa} ij} / n_{\alpha}$
    and
    (e-h) SH conductivity
    $\sigma_{\alpha} = \epsilon^{\alpha a}_{\phantom{\alpha a} ij} \sigma_{sa}^{\phantom{sa} ij} / n_{\alpha}$,
    in which $n_{\alpha} = (2, 4, 2, 2)$ are the numbers of nonzero components in $\epsilon_{\alpha a}^{\phantom{\alpha a} ij}$.
    Errors are defined by $\max_{\epsilon_{\alpha} \not= 0} |\gamma_{sa}^{\phantom{sa} ij} - \gamma_{\alpha}|$ etc.
    Black solid and orange dotted lines correspond to the number of iterations for wannierization $= 0$ and $1000$, respectively.
    Gray area represents the band gap.%
    } \label{fig:Figure2}
\end{figure*}

Figures~\ref{fig:Figure2}a-d show the Fermi-energy dependence of the SAC.
See Sec.~\ref{sub:Te_details} for computational details.
In this material, the $D_{3}$ point group symmetry allows the form of
\begin{equation}
    \gamma_{sa}^{\phantom{sa} ij}
    = \sum_{\alpha = 1}^{4} \gamma_{\alpha} \epsilon_{\alpha a}^{\phantom{\alpha a} ij}, \label{eq:spin_accumulation_d3}
\end{equation}
in which $\epsilon_{\alpha a}^{\phantom{\alpha a} ij}$ ($\alpha = 1, \dots, 4$) are tensors whose nonzero components are
$\epsilon_{1z}^{\phantom{1z} xy} = -\epsilon_{1z}^{\phantom{1z} yx} = 1$,
$\epsilon_{2x}^{\phantom{2x} xx} = -\epsilon_{2x}^{\phantom{2x} yy} = -\epsilon_{2y}^{\phantom{2y} xy} = -\epsilon_{2y}^{\phantom{2y} yx} = 1$,
$\epsilon_{3x}^{\phantom{3x} yz} = -\epsilon_{3y}^{\phantom{3y} xz} = 1$,
and $\epsilon_{4y}^{\phantom{4y} zx} = -\epsilon_{4x}^{\phantom{4x} zy} = 1$.
Our results are consistent with the symmetry and gauge invariant, which validates our implementation.

In Figs.~\ref{fig:Figure2}e-h,
we show the Fermi-energy dependences of the SH conductivities $\sigma_{\alpha}$ of the conventional spin current,
which are defined similarly to Eq.~\eqref{eq:spin_accumulation_d3}.
In the band gap painted in gray, we find $\sigma_{\alpha} = (-15, -2, -24, -4) (\hbar / e)~\mathrm{S / cm}$.
One of the critical problems in the SH conductivity is that it can be nonzero in insulators
even though the spin accumulation does not occur.
On the other hand, the SAC always vanishes in insulators because it is a Fermi-surface term as in Eq.~\eqref{eq:spin_accumulation}.
This is consistent with the fact that the spin accumulation is current-induced.

\section{Discussion}\label{sec:discussion}
To clarify the relevance of the SAC as an indicator of the SH effect,
let us summarize critical problems in the conventional spin current here.
First, the equilibrium expectation value can be nonzero in the absence of the inversion symmetry~\cite{PhysRevB.68.241315}.
This point is in a sharp contrast to the charge current,
whose equilibrium expectation value is forbidden by the Bloch-Bohm theorem~\cite{PhysRev.75.502}
even in the absence of the inversion and time-reversal symmetries.
Second, there is no conjugate force, and hence Onsager's reciprocity does not hold.
Finally, as pointed out above, the SH conductivity can be nonzero in insulators,
where the charge current does not flow and the spin accumulation is forbidden by the time-reversal symmetry.
The conventional spin current describes neither transport phenomena nor the spin accumulation.

The conserved spin current~\cite{PhysRevLett.96.076604} has some desirable properties.
First, the equilibrium expectation value takes the form of a magnetization current,
and hence the net equilibrium current vanishes,
if the spin torque quadrupole moment is considered~\cite{PhysRevB.104.L241411}.
Second, this current is conjugate to the Zeeman field gradient, and Onsager's reciprocity holds~\cite{PhysRevLett.96.076604}.
Even though this driving force realized in spin pumping experiments~\cite{1.2199473} has ac component only,
the first and second problems in the conventional spin current have been resolved.

These properties, however, are not sufficient to choose the conserved spin current.
The SH conductivity was recently computed for topological insulators with the help of maximally localized Wannier functions~\cite{Ma2024}.
The authors reported nonzero results, which were not ascribed to surface states, and hence the third problem is not resolved.
Note that the formula in ref.~\cite{Ma2024}, which was based on ref.~\cite{PhysRevB.108.195434},
does not take the spin torque quadrupole moment into account and differs from that in ref.~\cite{PhysRevB.104.L241411}.
More importantly, it has yet to be proved whether the conserved spin current, not any of the others, is experimentally observed,
in particular, via the damping-like spin torque in ferromagnetic resonance experiments~\cite{PhysRevLett.101.036601,PhysRevLett.106.036601}.

The SAC are free from the aforementioned problems.
First, it does not matter if the equilibrium expectation value of spin does not vanish.
Second, Onsager's reciprocity holds~\cite{PhysRevB.106.045203};
the inverse SH effect can be characterized by the response of the charge current to the time derivative of the Zeeman field gradient,
which has dc component in spin pumping experiments~\cite{1.2199473}, in contrast to the conserved spin current.
Third, the SAC itself vanishes in insulators as seen in Eq.~\eqref{eq:spin_accumulation}.
Furthermore, the exponential decay of the spin accumulation can be reproduced
if the diffusion propagator is taken into account~\cite{PhysRevB.98.174422}, as discussed more later.
The damping-like spin torque is also explained
by combination of the exchange coupling at the interface and the Gilbert damping~\cite{PhysRevB.106.045203}.
Thus, the SAC is a key parameter not only for the spin accumulation but also for recent SH experiments.

We also comment on two limitations of the SAC.
One is that we rely on the relaxation time approximation and neglect the vertex corrections.
Regarding the SH conductivity, the vertex corrections were taken into account in alloy systems
using the Korringa-Kohn-Rostoker method with the coherent potential approximation~\cite{PhysRevLett.106.056601}.
It is a future problem to evaluate the SAC based on the Green's functions~\cite{PhysRevB.106.045203} in the same manner.

The other is that we neglect the effect of diffusion.
Since the electric field gradient $\partial_{x^{i}} E_{j}$ has $\delta$-function peaks at the surface,
the induced spin density $\langle \Delta \hat{s}_{a} \rangle$ as well.
If we take the diffusion propagator into account, the response becomes nonlocal,
and the exponential decay can be reproduced~\cite{PhysRevB.98.174422}.
To see this, let us consider to apply a uniform electric field $E_{j0}$ to a finite section $-L / 2 < x^{i} < L / 2$.
In the reciprocal space, the electric field is expressed by $E_{j}(Q_{i}) = (2 E_{j0} / Q_{i}) \sin Q_{i} L / 2$.
Here we introduce the phenomenological diffusion factor $D_{s}(Q_{i}) = D_{s0} / [(Q_{i} l_{s})^{2} + 1]$,
and the response is expressed by
\begin{equation}
    \langle \Delta \hat{s}_{a} \rangle(Q_{i})
    = D_{s}(Q_{i}) g_{sa}^{\phantom{sa} ij} i Q_{i} E_{j}(Q_{i}). \label{eq:diffusion}
\end{equation}
$D_{s0}$ and the spin diffusion length $l_{s}$ depend on the details of disorder.
Back to the real space, the spin density turns out to decay exponentially as
\begin{align}
    \langle \Delta \hat{s}_{a} \rangle(x^{i})
    = & \frac{D_{s0} g_{sa}^{\phantom{sa} ij}}{2 l_{s}} E_{j0} \notag \\
    & \times (e^{-|x^{i} + L / 2| / l_{s}} - e^{-|x^{i} - L / 2| / l_{s}}). \label{eq:exponential}
\end{align}
Experimentally, the absoute magnitude of the spin distribution is available from the Kerr rotation spectroscopy; see ref.~\cite{Kato1910}.
Hence, the SAC can be experimentally measured using Eq.~\eqref{eq:exponential} with the mean free path and spin difussion length being measured.
Also, in a realistic setup such as ferromagnet--heavy-metal heterostructures,
the electric field gradient at the interface is unknown but likely small.
However, the spin accumulation occurs in response to the uniform electric field $E_{j0}$ as in Eq.~\eqref{eq:exponential}
and gives rise to the spin torque.

Note that we neglect the surface effects as the SAC is evaluated as a bulk property using Bloch wavefunctions.
Since the inversion symmetry is broken at the surface,
the resulting Rashba SOC may cause the additional spin relaxation
and the spin Edelstein effect~\cite{Ivchenko1978,Ivchenko1989,Aronov1989,Edelstein1990233}.
Also, in topological insulators that are gapped in the bulk~\cite{RevModPhys.82.3045},
the SAC vanishes, but the spin injection was experimentally succeeded~\cite{PhysRevLett.113.196601}.
This spin injection originates from the spin Edelstein effect owing to gapless surface states.
The SAC is not suitable for understanding the entire spin accumulation at the surface.
Our scope is the spin accumulation from the SH effect in the bulk.

We also neglect the intrinsic Fermi-sea term allowed in the absence of the time-reversal symmetry.
Although this term is related to the magnetic SH effect experimentally observed recently~\cite{Kimata2019},
its explicit formula using Bloch wavefunctions has not been derived yet.
Note that this term can be nonzero not only in metals but also in insulators as far as the system is magnetic.

To summarize, we have deloveped an \textit{ab initio} computational scheme for the SAC~\eqref{eq:spin_accumulation}
that characterizes the spin accumulation owing to the SH effect as a bulk property.
Using maximally localized Wannier functions, we can evaluate the SAC with dense $\bm{k}$-mesh for real materials.
We have applied our method to monolayer TMDC MoS$_{2}$ and trigonal tellurium
and checked the consistency with the point group symmetry and the gauge invariance as expected from the trace formula~\eqref{eq:trace_q}.
There is no theoretical justification for what definition of spin current, not any of the others, is experimentally observed.
We believe that the SAC is an alternative indicator of the SH effect without any ambiguity,
and its Wannier interpolation would pave the way to quantitative materials research on the SH effect.

\section{Methods}\label{sec:methods}
Our overall workflow is as follows.
First, we carry out \textit{ab initio} calculations with Vienna Ab initio Simulation Package (VASP)
based on the projector augmented wave (PAW) method~\cite{PhysRevB.54.11169,PhysRevB.59.1758}.
The SOC is taken into account.
Next, we construct Wannier functions with Wannier90~\cite{Pizzi_2020}.
We also generate additional input files \texttt{seedname.\{uHu, spn, sIu\}} from VASP output files via WannierBerri~\cite{Tsirkin2021}.
Finally, we compute the SAC and SH conductivity~\cite{PhysRevB.99.235113} using a post-process code of Wannier90, \texttt{postw90}.

\subsection{Computational details for monolayer MoS$_{2}$}\label{sub:MoS2_details}
Monolayer MoS$_{2}$, made from $2H$-MoS$_{2}$, belongs to the space group No.~$187$ ($D_{3h}^{1}$, $P\bar{6}m2$).
The lattice constants are not optimized from $2H$-MoS$_{2}$ in Materials Project~\cite{Jain2013} No.~$2815$
and set to $a =3.19~\mathrm{\AA}$ and $c = 13.38~\mathrm{\AA}$,
while the position of sulfur is optimized to the Wyckoff position $2h$ with $z = 0.116818$.
In order to deal with the vdW interaction,
we use rev-vdW-DF2 for the exchange correlation functional~\cite{PhysRevB.89.121103,PhysRevB.91.119902}.
We set the plain-wave energy cutoff to $258.689~\mathrm{eV}$ and $\bm{k}$-mesh to $48 \times 48 \times 2$.

\begin{figure}
    \centering
    \includegraphics[clip,width=0.5\textwidth]{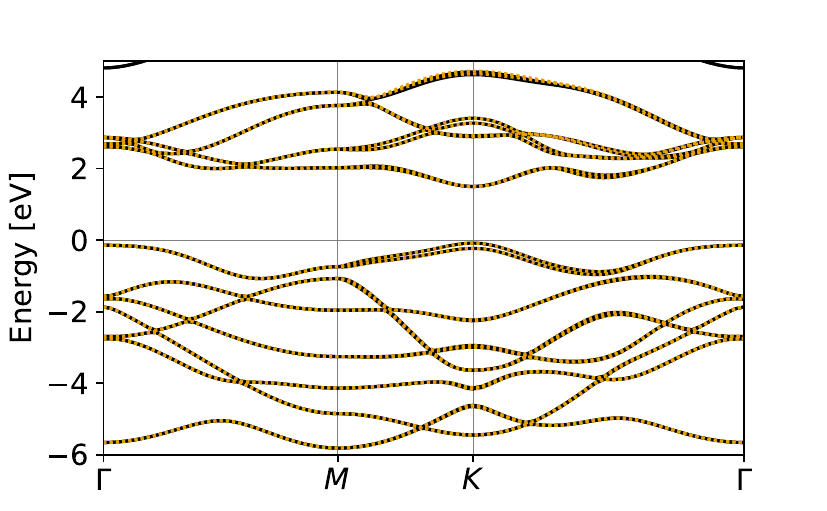}
    \caption{%
    Band structure of monolayer MoS$_{2}$ obtained by VASP (black solid line) and Wannier90 (orange square).%
    } \label{fig:Figure3}
\end{figure}

In Wannier90 calculations, we set the initial guess to $d$ orbitals of molybdenum and $p$ orbitals of sulfur,
which leads to $22$ Wannier functions,
$\bm{k}$-mesh to $12 \times 12 \times 1$,
the lower bound of the outer window to $-6~\mathrm{eV}$ measured from the Fermi energy,
the inner window to $[-6~\mathrm{eV}, 4~\mathrm{eV}]$,
and the number of iterations for wannierization to $0$ or $1000$.
The total Wannier spread is reduced from $36.157~\mathrm{\AA}^{2}$ to $35.881~\mathrm{\AA}^{2}$.
The obtained band structure is shown in Fig.~\ref{fig:Figure3}.

In \texttt{postw90} calculations, we set the smearing to $0.02~\mathrm{eV}$ and $\bm{k}$-mesh to $512 \times 512 \times 1$.
The SAC and SH conductivity per layer in Fig.~\ref{fig:Figure1} are obtained by multiplying the $c$-axis length.

\subsection{Computational details for trigonal tellurium}\label{sub:Te_details}
Trigonal tellurium is one of the most famous chiral materials belonging to the space group No.~$152$ ($D_{3}^{4}$, $P3_{1}21$).
We do not carry out structural optimization and
set the lattice constants to $a = 4.60~\mathrm{\AA}$ and $c = 5.90~\mathrm{\AA}$ in Materials Project~\cite{Jain2013} No.~$19$.
Tellurium is located at the Wyckoff position $3a$ with $x = 0.256697$.
We use the generalized gradient approximation proposed by Perdew, Burke, and Ernzerhof
for the exchange correlation functional~\cite{PhysRevLett.77.3865,PhysRevLett.78.1396}.
We set the plain-wave energy cutoff to $174.982~\mathrm{eV}$ and $\bm{k}$-mesh to $24 \times 24 \times 24$.

\begin{figure}
    \centering
    \includegraphics[clip,width=0.5\textwidth]{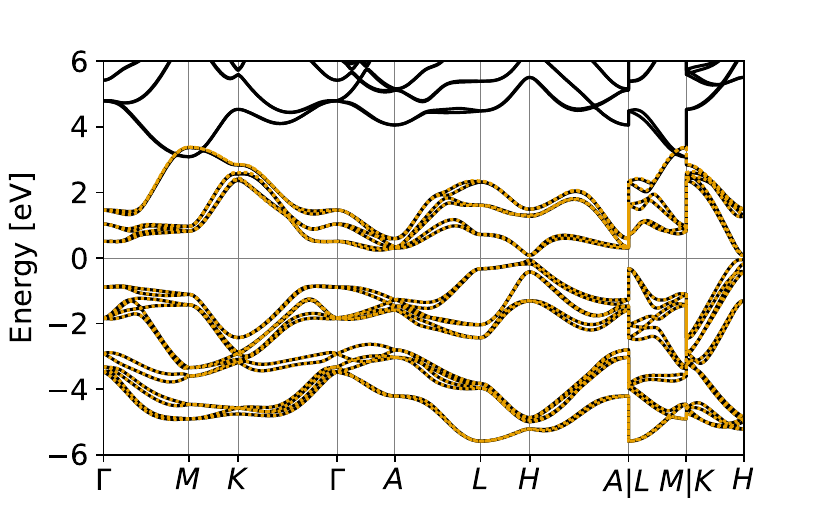}
    \caption{%
    Band structure of trigonal tellurium obtained by VASP (black solid line) and Wannier90 (orange square).%
    } \label{fig:Figure4}
\end{figure}

In Wannier90 calculations, we set the initial guess to $p$ orbitals, which leads to $18$ Wannier functions,
$\bm{k}$-mesh to $6 \times 6 \times 6$,
the inner window to $[-6~\mathrm{eV}, 2~\mathrm{eV}]$ measured from the Fermi energy,
and the number of iterations for wannierization to $0$ or $1000$.
The total Wannier spread is reduced from $50.186~\mathrm{\AA}^{2}$ to $50.048~\mathrm{\AA}^{2}$.
The obtained band structure is shown in Fig.~\ref{fig:Figure4}.

In \texttt{postw90} calculations, we set the smearing to $0.02~\mathrm{eV}$ and $\bm{k}$-mesh to $256 \times 256 \times 256$.

\section*{Data availability}
The data generated in this work will be made available upon reasonable request to the first author.

\section*{Code availability}
The code used to generate the results in this work will be made available upon reasonable request to the first author.

%

\begin{acknowledgments}
    We thank R.~Arita and T.~Koretsune for their advice on Wannier90, and K.~Kondo and Y.~Niimi on SH experiments.
    This work was supported by the Japan Society for the Promotion of Science KAKENHI (Grants No.~JP22K03498 and No.~JP23K21081).
\end{acknowledgments}

\section*{Author contributions}
A.S. developed the theory, implemented the code, and carried out calculations.
A.S. and E.M. wrote the manuscript.

\section*{Competing interests}
The authors declare no competing interests.
\end{document}